\pdfoutput=1

\documentclass[reprint,amsmath,amssymb,floatfix,aps,longbibliography]{revtex4-1}
\usepackage{graphics,graphicx,float}

\usepackage[pdftex,
            pdfauthor={Steven A. Frank},
            pdftitle={},
            pdfsubject={Demography and the tragedy of the commons}]{hyperref} 

\def\cite{\citep}
\def\shortcite{\citep}
\def\citeA{\textcite}

\newcommand{\bx}{\mbox{\bf x}}
\newcommand{\by}{\mbox{\bf y}}
\newcommand{\bz}{\mbox{\bf z}}

\newcommand{\Ga}{\alpha}

\newcommand{\Gd}{\delta}

\newcommand{\Gg}{\gamma}

\newcommand{\Gl}{\lambda}

\newcommand{\Gs}{\sigma}

\newcommand{\Gth}{\theta}

\DeclareMathOperator{\E}{E}
\newcommand{\dd}{{\hbox{\rm d}}}

\newcommand{\Eq}[1]{Eq.~(\ref{eq:#1})}

\newcommand{\ovr}[2]{{{#1}\over{#2}}}
\newcommand{\dovr}[2]{\ovr{\dd #1}{\dd #2}}

\newcommand{\prt}{\partial}
\newcommand{\povr}[2]{\ovr{\prt #1}{\prt #2}}

\begin{document}

\title{Demography and the tragedy of the commons}

\author{Steven A.\ Frank}
\email[email: ]{safrank@uci.edu}
\homepage[homepage: ]{http://stevefrank.org}
\affiliation{Department of Ecology and Evolutionary Biology, University of California, Irvine, CA 92697--2525  USA}

\begin{abstract}

Individual success in group-structured populations has two components.  First, an individual gains by outcompeting its neighbors for local resources.  Second, an individual's share of group success must be weighted by the total productivity of the group.  The essence of sociality arises from the tension between selfish gains against neighbors and the associated loss that selfishness imposes by degrading the efficiency of the group.  Without some force to modulate selfishness, the natural tendencies of self interest typically degrade group performance to the detriment of all.  This is the tragedy of the commons.  Kin selection provides the most widely discussed way in which the tragedy is overcome in biology.  Kin selection arises from behavioral associations within groups caused either by genetical kinship or by other processes that correlate the behaviors of group members.  Here, I emphasize demography as a second factor that may also modulate the tragedy of the commons and favor cooperative integration of groups.  Each act of selfishness or cooperation in a group often influences group survival and fecundity over many subsequent generations.  For example, a cooperative act early in the growth cycle of a colony may enhance the future size and survival of the colony.  This time-dependent benefit can greatly increase the degree of cooperation favored by natural selection, providing another way in which to overcome the tragedy of the commons and enhance the integration of group behavior.  I conclude that analyses of sociality must account for both the behavioral associations of kin selection theory and the demographic consequences  of life history theory\footnote{\href{http://dx.doi.org/10.1111/j.1420-9101.2009.01893.x}{doi:\ 10.1111/j.1420-9101.2009.01893.x} in \textit{J. Evol. Biol.}}

\end{abstract}

\maketitle

\begin{quote}
In a single battle the Peloponnesians and their allies may be able to defy all Hellas, but they are incapacitated from carrying on a war\ldots by the want of the single council-chamber requisite to prompt and vigorous action\ldots Slow in assembling, they devote a very small fraction of the time to the consideration of any public object, most of it to the prosecution of their own objects. Meanwhile each fancies that no harm will come of his neglect, that it is the business of somebody else to look after this or that for him; and so, by the same notion being entertained by all separately, the common cause imperceptibly decays \cite[pp.~93--94]{thucydides14history}.
\end{quote}

\section*{Introduction}

Thucydides describes how self interest works against group benefit and ultimately degrades individual success.  \citeA{hardin68tragedy} coined the phrase ``the tragedy of the commons'' for this process.  The tragedy shapes all patterns of life because of the fundamental self interest promoted by natural selection. 

In evolutionary studies, the central role of the tragedy grew with increasing understanding of kin selection, group selection, selfish genes, and levels of selection \shortcite{rankin07the-tragedy}. \citeA{hamilton64the-genetical,hamilton67extraordinary} had the idea from the start, and the problem informed much of his great work.  The vigorous debates in the 1980s over sex ratios and levels of selection turned on how population structure and genetic relatedness modulated the tragedy \cite{wilson81evolution,grafen84natural,nunney85female-biased,frank86hierarchical,frank06social}.  Later, Maynard Smith and Sz{\'a}thmary began to emphasize that the major transitions in the history of life followed solution of the tragedy at various levels of organization: the integration of genomes, cells, multicell organisms, and colonies into functional units with relatively little internal conflict \cite{szathmary87group,maynard-smith88evolutionary,szathmary89the-emergence,maynard-smith95the-major}.

\citeA{leigh77how-does} directly connected the puzzles of group integrity to Hardin's ``tragedy of the commons'' slogan, but, at the time, that framing did not catch on \cite{rankin07the-tragedy}.  In addition, none of the work had laid out the tragedy in a simple and explicit evolutionary model and connected the ideas to the broader subject raised by Hardin.  I picked up on this opportunity and made several explicit evolutionary models of the tragedy that directly linked the concepts to Hardin's memorable phrase \cite{frank94kin-selection,frank95mutual,frank96models,frank96policing}.

I showed the interaction of two key ideas.  First, behavioral association through genetic relatedness provides the simplest solution to the tragedy.  In the tradeoff between individual success and group efficiency, natural selection favors more closely related individuals to act more like a unit.  The group achieves improved efficiency in proportion to the relatedness between group members. My model showed this well-known relatedness effect of group population structure in the simplest way, facilitating application to a wide variety of problems.  

The second idea followed the work of \citeA{leigh77how-does}, \citeA{alexander79darwinism,alexander87the-biology}, and \citeA{buss87the-evolution}, in which they emphasized mechanisms that prevented or repressed internal competition within groups.  If individuals cannot compete within groups, they can only increase their own personal success by raising the efficiency of the group in which they live \cite{frank03repression}.  The evolutionary problem concerns how such internal repressive mechanisms can be favored by natural selection.  Any individual that pays a cost toward such a mechanism would be at a competitive disadvantage against neighbors that refrain from contributing to the group-beneficial process.  I showed, in an explicit evolutionary model, how such repressive mechanisms can arise, and how the dynamics depend on an interesting interaction with kin selection \cite{frank95mutual}.

Since the mid-1990s, the tragedy of the commons has become the central concept in much work on sociality \cite{dionisio2006tragedy,rankin07the-tragedy}.  On the theoretical side, the rapid growth of the slogan and the approach derives from greatly increased interest in the evolutionary analysis of cooperation.  On the empirical side, more has become known about various levels of social integration and conflict in insects and other animals. 

Several new observations have also been made on social aspects of microbes \shortcite{crespi01the-evolution,west07the-social,nadell09the-sociobiology}.  For example, many microbes secrete molecules to alter the environment in beneficial ways.  The secretions are often costly to the secretors with regard to growth rate.  Neighbors that do not secrete gain the same benefit as the secretors, but do not pay the cost.  This creates a tragedy of the commons: all microbes in a group do better by secretions that modify the environment, but those cheaters that do not secrete can outcompete their cooperative, secreting neighbors.  As the cheaters rise in frequency, group efficiency declines.

Recent studies of the tragedy focus almost entirely on the role of behavioral associations and the kin selection coefficient of relatedness in determining the level of cooperation and group efficiency.  Here, I argue that demographic factors may often be as important as relatedness in shaping the level of cooperative behavior and the degree to which groups succumb to the tragedy.  

\section*{The tragedy and behavioral associations}

I first establish the basic model of the tragedy with respect to behavioral associations expressed by the coefficient of relatedness of kin selection theory.  I then extend the model to analyze the consequences of demography.

The role of relatedness can be understood simply by Hamilton's \citeyear{hamilton64the-genetical} rule, in which a cooperative behavior increases if $rb-c>0$.  In group-structured models, the term $r$ measures the behavioral association between group members \cite{frank98foundations}.  For models of the tragedy of the commons, typically all members of a local group are both actors and recipients with regard to cooperative behavior.  A pathogen may, for example, secrete a molecule that interferes with host immunity.  All members of the local pathogen population, including the secretor itself, gain by the secretion.  Thus we may think of the group as the recipient and of each individual as a potential secretor and thus as an actor.

The terms $b$ and $c$ represent the benefit to the recipients and the cost to the actor.  If we know $r$, $b$, and $c$, then we can easily evaluate how a cooperative trait evolves.  However, the ways in which $b$ and $c$ arise in relation to behaviors can be complex.  

Suppose, for example, that we follow a pathogen population within a host over the course of an infection.  Let time $t_0$ be the initiation of the infection.  At time $t_1$, individuals may secrete a quantity $1-x$ of a molecule that benefits all members of the group. Suppose that the competitiveness of an individual that secretes $1-x$ is in proportion to $x$; for example, a secretion of $1-x=0$ allows the individual to achieve the maximum relative competitiveness of $x=1$.  We can think of $x$ as the level of competitiveness against neighbors in the group and, equivalently, we can think of $1-x$ as the level of cooperation.  Define the average level of cooperation in the local group at time $t_1$ to be $1-y$; thus, the average level of competitiveness is $y$.

Assume that the cooperation level at time $t_1$ has only the following two immediate consequences.  First, relative competitiveness of an individual compared with neighbors is in proportion to $x/y$.  This fraction is proportional to the relative share of group productivity obtained by an individual that cooperates at level $1-x$ in a group with average cooperation level $1-y$.  Second, cooperation enhances group productivity.  Thus, group productivity is in proportion to the average group cooperation, $1-y$.  

Multiplying an individual's relative share of group productivity, $x/y$, by the group productivity, $1-y$, yields $\E(w|x,y)$, the expected fitness of an individual given $x$ and $y$ \cite{frank94kin-selection}, as
\begin{equation}\label{eq:wStandard}
	w=\ovr{x}{y}(1-y),
\end{equation}
where I write $w$ as a shortened notation for $\E(w|x,y)$.  We find the phenotype favored by natural selection as the value of competitiveness, $z^*$, and the level of cooperation, $1-z^*$, by using the method introduced in \citeA{frank95mutual} and developed in \citeA{taylor96how-to-make}.  To keep the analysis simple and focus on essential processes, I assume here that individuals are haploid.  

We obtain the favored phenotype by maximizing $w$ with respect to $x$, and evaluating at a candidate equilibrium $x=y=z^*$.  In particular, we solve
\begin{align}
	\dovr{w}{x}	&= \povr{w}{x} + \povr{w}{y}\dovr{y}{x}\label{eq:method}\\
				&= -c + br = 0.\notag
\end{align}
From Hamilton's rule, the equilibrium occurs when $rb-c=0$.  Hamilton's rule was known since 1964, but was not used successfully to solve problems of sex ratios, dispersal, or tragedy of the commons interactions \cite{frank98foundations}.  The method here shows how to go directly from an expression for fitness, $w$, to a solution, by calculating the marginal cost as $-c=\prt w/\prt x$, the marginal benefit as $b=\prt w/\prt y$, and relatedness as $r=\dd y/\dd x$.  

Applying this method to \Eq{wStandard} and evaluating at $x=y=z^*$ yields
\begin{align*}
	\dovr{w}{x}	&= \povr{w}{x} + \povr{w}{y}\dovr{y}{x}\\
				&= \ovr{1-z^*}{z^*} - r\left(\ovr{1}{z^*}\right) = 0.
\end{align*}
Solving gives the level of secretion and the degree of cooperation \cite{frank94kin-selection,frank95mutual} as
	\[1-z^*=r,\]
which we may also express as the ratio of cooperative to competitive tendency, $1-z^*:z^*$, as
\begin{equation}\label{eq:ratio1}
	r:1-r.
\end{equation}
The method expressed in \Eq{method} \cite{taylor96how-to-make,frank97multivariate} has been used to solve many problems.  Before 1996, I used variants of this method to analyze numerous models of dispersal, sex ratios, and social evolution \cite{frank98foundations}.  Indeed, most such problems had not been, and effectively could not have been, solved simply in general terms of behavioral associations and relatedness without this sort of method.  Since 1996, many others have picked up the method and applied it to a wide variety of problems in social evolution.

The method of \Eq{method} provides a simple way to parse the components of Hamilton's rule.  This method is one contribution of \citeA{taylor96how-to-make} and, as I mentioned, this method has been widely used in recent years.  The second contribution of \citeA{taylor96how-to-make} is a  formal method to combine demographic analyses of life history with the social aspects that derive from the kin selection coefficient of relatedness.  That approach arose from combining Taylor's \citeyear{taylor90allele} formal methods for life history analysis with the simple parsing of components of fitness achieved by an extension of \Eq{method}.  

The combination of life history theory with kin selection theory provided, for the first time, a simple and direct way to study the relative importance of components of fitness with regard to social traits.  In particular, the method allowed one to understand how a social behavior may simultaneously influence aspects of current and future fecundity and survival, with those distinct fitness components weighted by the various kinds of phenotypic and genetic correlations between different classes of individuals that influence the selection and transmission of social characters.  

The life history components of fitness often shape social traits as strongly as do the genetic associations emphasized in the kin and group selection theories of sociality.  Much of my book on social evolution emphasized how demography shapes life history and sociality \cite{frank98foundations}.  However, in subsequent years, those demographic and life history factors have not received nearly the attention as the kin and group selection factors of sociality.

The tragedy of the commons model has become the archetypical problem of social evolution.  To show the ways in which demography and life history contribute to sociality, I extend the typical expression of the tragedy model given above to include explicitly the fecundity and viability components of fitness. 

\section*{The tragedy in a demographic context}

To illustrate the relative contributions of demography and relatedness to cooperation, I extend the tragedy of the commons model.  I use a variant of a sex ratio model in \citeA[pp.~238--242]{frank87demography,frank98foundations}. Several other sex ratios models addressed related problems \shortcite{bulmer80sex-ratio,wilson81evolution,frank86hierarchical,tienderen86sex-ratio,aviles93interdemic,reece05host,nagelkerke96hierarchical}.  See also various other formulations of demography and cooperation \shortcite{kokko01the-evolution,gardner07is-bacterial,lehmann07the-evolution,rankin07resolving,alizon08empty,johnstone08sex-differences,lion09habitat}.

\subsection*{Cycle fitness}

We follow a group or colony through time.  For simplicity, I focus on a series of discrete generations indexed by $j=0,\ldots,\infty$.  Consider, in each generation, the competitiveness of an individual, $x_j$, or, equivalently, the individual's level of cooperation, $1-x_j$.  Similarly, we use group competitiveness, $y_j$, or group cooperation, $1-y_j$.  Individuals may have different behaviors in each generation, $j$, altering their competitiveness, $x_j$.  We seek the set of behaviors favored by natural selection, $\bz^*=\{z_j^*\}$ for $j=0,\ldots,\infty$.

To start, we need an expression for how an individual's fitness depends on its own behavior, $\bx=\{x_j\}$, and the average behavior of the individual's group, $\by=\{y_j\}$.

The fitness consequences of individual and colony behavior, $x_j$ and $y_j$, can be expressed by three factors measured over a full demographic cycle \cite[p.~239]{charlesworth80evolution,frank98foundations}.  First, the individual and colony behaviors in generation $j$ determine the fraction, $f$, of future progeny that descend from a generation $j$ individual. Second, future progeny must be discounted by the population growth rate, $\Gl$. Finally, the number of future progeny depends on the fecundity of the colony, $F$, at each age multiplied by the probability that the colony will survive to that age, $S$. These factors combine to give the total reproductive value of future progeny that emigrate to form new generation 0 colonies 
\begin{equation}\label{eq:wDemographic}
	w_j = f(x_j,y_j)\sum_{k=j}^{\infty}\Gl^{-k}S(\by_k)F(\by_k).
\end{equation}
The cycle fitness of an individual in generation $j$, with competitiveness $x_j$ and colony competitiveness $y_j$, is given by $w_j$. The function $f$ is the fraction of future colony offspring that descend from an individual with competitiveness $x_j$. The survivorship of the colony to produce generation $k$ is $S(\by_k)$, where $\by_k$ is the vector of all colony competitiveness values for $j = 0,\ldots, k$. The fecundity of the colony in generation $k$ is $F(\by_k)$.

\subsection*{General solution}

If the competitiveness produced in each generation is an independent trait, then the direction of selection on competitiveness in the $j$th generation in this haploid model is given by $\dd w_j/\dd x_j$. Because there are three functions, $f$, $S$, and $F$, the total derivative has three parts by application of the chain rule
\begin{equation}\label{eq:Psum}
	\dovr{w_j}{x_j} = P_1+P_2+P_3.
\end{equation}
We search for a candidate solution by solving 
\begin{equation}\label{eq:PsumZero}
	\dovr{w_j}{x_j}=0
\end{equation}
at $\bx=\by=\bz^*$.

Before expressing the individual components of \Eq{Psum}, it is useful to have some shortened notation, in which all derivatives are evaluated at $\bx=\by=\bz^*$
\begin{align*}
	K &= \sum_{k=j}^{\infty}\Gl^{-k}S(\bz_k^*)F(\bz_k^*)\\
	r &= \dd y_j/\dd x_j\\
	f^* &= f(z_j^*,z_j^*)\\
	f_{x_j} &= \povr{f(x_j,y_j)}{x_j}\\
	f_{y_j} &= \povr{f(x_j,y_j)}{y_j}\\
	F_{y_j} &= \ovr{\prt F(\by_k)}{F(\bz_k^*)\prt y_j}
						= \povr{\log\left[F(\by_k)\right]}{y_j}\\
	S_{y_j} &= \ovr{\prt S(\by_k)}{S(\bz_k^*)\prt y_j}
						= \povr{\log\left[S(\by_k)\right]}{y_j}.
\end{align*}

Differentiating $f$, $S$, and $F$ in turn, in each case holding the other two terms constant, yields
\begin{align*}
	P_1 &= \left(f_{x_j}+rf_{y_j}\right)K\\
	P_2 &= rf^*\sum_{k=j}^{\infty}\Gl^{-k}S(\bz_k^*)F(\bz_k^*)S_{y_j}\\
	P_3 &= rf^*\sum_{k=j}^{\infty}\Gl^{-k}S(\bz_k^*)F(\bz_k^*)F_{y_j}.
\end{align*}
If $S_{y_j}$ and $F_{y_j}$ are independent of the index $k$, as in the specific model below, then we can write these expressions in simplified form
\begin{align*}
	P_1 &= \left(f_{x_j}+rf_{y_j}\right)K\\
	P_2 &= rf^*KS_{y_j}\\
	P_3 &= rf^*KF_{y_j}.
\end{align*}

We obtain the behavior in the $j$th generation favored by natural selection, $z_j^*$, by solving \Eq{PsumZero}.  Using the simplified forms for the $P$ terms and substituting into \Eq{PsumZero} yields the expression
\begin{equation}\label{eq:soln}
	f_{x_j} + rf_{y_j}+rf^*\left(S_{y_j} + F_{y_j}\right) = 0.
\end{equation}
With respect to Hamilton's rule, $rb-c=0$ at equilibrium. Here, we have $-c=f_{x_j}$ and $b=f_{y_j}+f^*\left(S_{y_j} + F_{y_j}\right)$.  The survival and fecundity components, $S$ and $F$, show explicitly the marginal effects of altruism associated with these two components of fitness.  In some cases, a current behavior may cause marginal changes in survival or fecundity in the future.  This formulation accounts for those future effects.

\subsection*{Simple tragedy model}

I extend the approach to the simple tragedy model in \Eq{wStandard} to a multigenerational model in which we need to track survival and fecundity.  To begin, let individual competitiveness against neighbors lead to a relative success in the $j$th generation of 
	\[f(x_j,y_j)=\ovr{x_j}{y_j}.\]
Then $f^*=1$ and 
	\[f_{x_j} + rf_{y_j}=\ovr{1-r}{z_j^*}.\]

We now need to make some assumptions about how competitive and cooperative behaviors influence survival and fecundity.  In this particular model, I assume that the colony grows from generations $k=0,\ldots,g-1$ without sending out any migrants, and then maintains a stable size and sends out migrants in proportion to $F(z_k^*)$ in the following generations $k=g,\ldots$ In this example, colony fecundity increases linearly with colony size; that is, the number of neighboring individuals neither increases nor decreases the fecundity per individual. Thus colony fecundity is zero through the first $g-1$ generations. Colony fecundity in the following generations, $k\ge g$, is proportional to
\begin{equation*}
	F(z_k^*)=\left[N\prod_{i=0}^{g-1}n(1-z_i^*)\right]n(1-z_k^*),
\end{equation*}
where $N$ is the number of founding individuals in generation 0, and $n(1 - z_i^*)$ is the number of offspring in the $i$th generation of a normal colony. The term in square brackets is the size that the colony has achieved during the growth phase, and $1 - z_k^*$ is the number of individuals produced for dispersal during the reproductive phase. When the level of cooperation deviates from normal only in generation $j$, then
\begin{equation*}
	F(\by_k)=F(\bz_k^*)\ovr{1-y_j}{1-z_j^*},
\end{equation*}
and the partial derivative of the logarithm of $F$ with respect to the deviant group level of cooperation is
\begin{equation*}
	F_{y_j}=-\ovr{1}{1-z_j^*}.
\end{equation*}

Colony survival is also divided into two periods. For colony growth, during generation $k < g$, survival in each generation is a function of the current colony size relative to the size of a mature, normal colony
\begin{equation*}
	\Gs(\by_k)=\Gd\left[\ovr{\prod_{i=0}^{k-1}n(1-y_i)}{\prod_{i=0}^{g-1}n(1-z_i^*)}\right]^\Gth,
\end{equation*}
with $\Gs(\by_k) = \Gd$ for $k\ge g$. Survival through generation $k$ is therefore 
\begin{equation*}
	S(\by_k)=\prod_{i=0}^k\Gs(\by_i).
\end{equation*}
When the level of cooperation deviates from normal only in generation $j < g - 1$, then survival in each generation with $k > j$ is 
\begin{equation*}
	\Gs(\by_k)=\Gs(\bz_k^*)\left[\ovr{1-y_j}{1-z_j^*}\right]^\Gth,
\end{equation*}
and survival in generations $k \le j$ is $\Gs (\bz_k^*)$. Cumulative survival over generations for $k > g - 1$ is
\begin{equation*}
	S(\by_k)=S(\bz_k^*)\left[\ovr{1-y_j}{1-z_j^*}\right]^{\Gth(g-1-j)}.
\end{equation*}
The partial derivative of the logarithm of $S$ with respect to deviant group cooperation is
\begin{equation*}
	S_{y_j}=-\ovr{\Gg_j}{1-z_j^*},
\end{equation*}
where
\begin{equation}\label{eq:gamma}
	\Gg_j = 
		\begin{cases}
			\Gth(g-1-j) & j < g-1 \\
			0 & j \ge g-1.
		\end{cases}
\end{equation}

We use these various expressions in \Eq{soln} to solve for the equilibrium level of cooperation in each generation $j$ as
\begin{equation}\label{eq:mainResult}
	1-z_j^* = r + \Ga_j,
\end{equation}
where
\begin{equation}\label{eq:alpha}
	\Ga_j = \ovr{r\Gg_j(1-r)}{1+r\Gg_j}
\end{equation}
is the extra amount of cooperation favored by the contribution of cooperation to colony survival.  We may also express the ratio of cooperative to competitive tendency, $1-z_j^*:z_j^*$, as
\begin{equation}\label{eq:ratio2}
	r(1+\Gg_j) : 1 - r,
\end{equation}
which, in comparison with \Eq{ratio1}, shows the simple consequence of the life history component summarized by $\Gg_j$.

\section*{Discussion}

The tragedy of the commons is the primary concept by which we understand competition and cooperation within groups.  Recent literature has emphasized the role of kin selection in modulating the level of cooperation favored by natural selection.  

I emphasized here that, in determining the level of cooperation within groups, the consequences of behavior for group survival and fecundity are just as important as the behavioral associations of kin selection.  To understand the tragedy, or any sort of sociality in groups and colonies, one must study natural selection within the full life history context of how behaviors influence survival, fecundity, and dispersal to form new colonies.  

\subsection*{Summary of the models}

The first model, in \Eq{wStandard}, isolated the role of kin selection to show clearly the direct effect of this important component.  In that model, competition against neighbors determines the share of group productivity acquired by an individual.  The total group productivity declines in proportion to the average level of competitiveness of group members---or, equivalently, group productivity rises in proportion to the average level of cooperation in the group.  The level of cooperation favored by natural selection is
	\[1-z^*=r,\]
where $r$ is the behavioral association between group members, usually taken as the coefficient of relatedness from kin selection theory.  

\begin{figure*}[t]
\includegraphics[width=0.85\hsize]{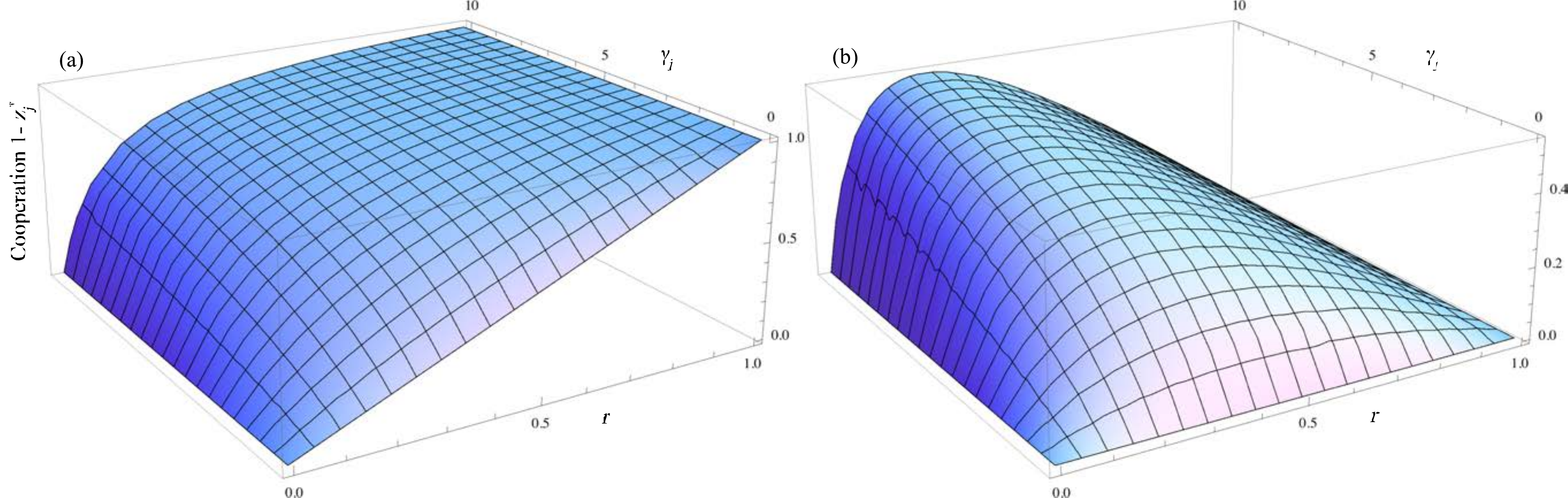}
\caption{Level of cooperation from relatedness, $r$, and colony survival benefits, $\Gg_j$, in the $j$th generation of colony development.  (a) The total level of cooperation combining the direct contribution of relatedness, $r$, plus the synergism between relatedness and survival given by $\Ga_j$ in \Eq{mainResult}. (b) The part of the total cooperation level ascribed to $\Ga_j$, the synergism between relatedness and survival.  The term $\Ga_j$ is defined in \Eq{alpha}.}
\end{figure*}

The variance between groups is in proportion to $1-r$, so we see directly the formal equivalence of kin selection theory based on associations within groups, $r$, and group selection theory, based on variance between groups, $1-r$ \cite{frank86hierarchical}.  Some authors object to the word ``kin'' in kin selection theory.  But, in the formal theory, it has long been known that the $r$ of kin selection describes the association between individuals in behavior or in genetic effects and not the pedigree relations usually associated with the word ``kin'' \cite{hamilton70selfish,hamilton75innate,frank98foundations}.  Instead of ``kin selection theory'' we could say ``behavioral association theory'', but the former has clear precedence.  

The second model, in \Eq{wDemographic}, takes account of the fact that behaviors in the present have future consequences for survival and reproduction.  If a young colony rarely reproduces when small, then a major effect of competition and cooperation in small colonies arises through the future consequences of those behaviors on colony survival and reproduction.  To illustrate these life history consequences, I made some simple assumptions about how behaviors affect survival and reproduction.  

In particular, I assumed that colonies first grow to a fixed size over a fixed period of time before reproducing \cite{oster78caste}.  Then, after the mature colony size has been reached, I assumed that colonies remain constant in size and export all extra growth as dispersers to found new colonies.  Throughout the colony lifetime, colony survival increases with colony size by setting the probability of survival in each generation in proportion to colony size raised to the exponent $\Gth$.  

Within each generation of the colony, I assumed that an individual's share of the total colony in the next generation is in proportion to the individual's competitiveness.  I also assumed that colony growth declines in proportion to the average level of competitiveness of group members---or, equivalently, group productivity rises in proportion to the average level of cooperation in the group.  

With these assumptions, I derived the level of cooperation in the $j$th generation of colony growth in \Eq{mainResult}, which I repeat here
	\[1-z_j^* = r + \Ga_j,\]
where $r$ is the behavioral association between individuals within groups, and $\Ga_j$ is the excess level of cooperation favored by the synergism between the behavioral association and the total future colony survival benefit added by an increased level of cooperation.  \Eq{alpha} gives the expression for $\Ga_j$, and \Eq{gamma} gives the expression for the future colony survival benefit from increased cooperation, $\Gg_j$.

Figure 1a shows the level of cooperation favored for each level of behavioral association, $r$, and colony survival benefit, $\Gg_j$.  The common interpretation of the tragedy solely in terms of relatedness, $r$, is given by $\Gg_j=\Ga_j=0$.  Figure 1b shows the extra amount of cooperation from the synergism between the behavioral association, $r$, and the life history component of survival benefits, $\Gg_j$.  The plots show that, at lower levels of relatedness, $r$, high levels of cooperation are still favored when the future survival benefits of cooperation are strong.  Thus, weakly related groups may still be highly cooperative and integrated when their mutual survival over time depends on strongly cooperative contributions from group members.  

In this particular model, I have allowed individuals to adjust their level of cooperation over time.   Early in the colony life cycle, cooperation strongly enhances future survival because colony survival depends on colony growth.  Later in the colony life cycle, the colony will have achieved its maximal size, and cooperation no longer enhances future survival.  Consequently, cooperation is more strongly favored early in the colony life cycle, with the measure of survival benefits for cooperation, $\Gg_j$, starting high when generation $j$ is low and declining to zero as generation $j$ increases.

\subsection*{Alternative assumptions}

The particular patterns of cooperation depend on the specific assumptions about how cooperation influences survival and fecundity.  For example, behavior could be fixed and unchanging with changes in colony size.  Then, as shown in \citeA{frank87demography} for a sex ratio model, the level of cooperation will be some sort of averaging over the cooperative intensity favored in each separate generation.  For example, in the specific model above, higher cooperation is favored early in the colony life cycle  and lower cooperation is favored later in the cycle.  With fixed behavior, the level of cooperation would be an averaging over the various levels favored over the colony life cycle.  

Alternatively, we might assume that cooperation does not affect colony survival in the early generations, when the colony is small, but cooperation can have a very strong effect on survival once the colony has achieved a certain size.  Such size dependence may arise because cooperation is not effective at small colony size.  For example, if cooperation occurs in a bacterial population through secretion of a diffusible molecule, then a small population may not be able to make enough of the diffusible molecule to alter the environment in a significant way.  As the colony grows, it eventually achieves sufficient density for the cooperative effects of the diffusible molecule to be significant.  In this case, cooperation likely rises as the colony grows, with the level of cooperation depending on how changes in behavior alter marginal survival and fecundity.  

The analyses here all depend on the assumption that natural selection favors those behaviors that contribute most to the future of the aggregate population.  There can, however, be strong components of natural selection operating on various timescales.  For example, mutants can arise within groups and spread rapidly, even though those mutants, by degrading the group in which they live, contribute little to the future population.  This tension between short and long timescales can be particularly strong in large, multigenerational microbial colonies, in which there is much opportunity for mutation and selection within groups \cite{levin94short-sighted}.  These timescale issues are very important, but distinctive with regard to methods of analysis and consequences.  I take up these issues in a later paper.

In summary, analyses of sociality must account for both the behavioral associations of kin selection theory and the demographic consequences  of life history theory.

\section*{Acknowledgments}

Parts of this paper were modified from \citeA[pp.~238--242]{frank87demography,frank98foundations}. My research is supported by National Science Foundation grant EF-0822399, National Institute of General Medical Sciences MIDAS Program grant U01-GM-76499, and a grant from the James S.~McDonnell Foundation.  

\bibliography{microbe}

\end{document}